\documentclass[aip,reprint]{revtex4-1}

\draft 

\usepackage{graphicx}
\usepackage{dcolumn}
\usepackage{bm}

\begin{document}


\title{Integrated spiral waveguide amplifiers on erbium-doped thin-film lithium niobate} 



\author{Xiongshuo Yan$ ^{\ast} $}
\affiliation{State Key Laboratory of Advanced Optical Communication Systems and Networks, School of Physics and Astronomy, Shanghai Jiao Tong University, Shanghai 200240, China}

\author{Yi'an Liu$ ^{\ast} $}
\affiliation{State Key Laboratory of Advanced Optical Communication Systems and Networks, School of Physics and Astronomy, Shanghai Jiao Tong University, Shanghai 200240, China}

\author{Jiangwei Wu}
\affiliation{State Key Laboratory of Advanced Optical Communication Systems and Networks, School of Physics and Astronomy, Shanghai Jiao Tong University, Shanghai 200240, China}

\author{Yuping Chen}
\email{ypchen@sjtu.edu.cn}
\affiliation{State Key Laboratory of Advanced Optical Communication Systems and Networks, School of Physics and Astronomy, Shanghai Jiao Tong University, Shanghai 200240, China}

\author{Xianfeng Chen}
\email{xfchen@sjtu.edu.cn}
\affiliation{State Key Laboratory of Advanced Optical Communication Systems and Networks, School of Physics and Astronomy, Shanghai Jiao Tong University, Shanghai 200240, China}



\date{\today}

\begin{abstract}
Integrated optical amplifiers and light sources are of great significance for photonic integrated circuits (PICs) and have attracted many research interests. Doping rare-earth ions in materials as a solution to realize efficient optical amplifiers and lasing has been investigated a lot. 
We investigate the erbium-doped lithium niobate on insulator (LNOI).
Here, spiral waveguide amplifiers were fabricated on a 1-mol\% erbium-doped LNOI by CMOS-compatible technique. We demonstrated a maximum internal net gain of 8.3 dB at 1530 nm indicating a net gain per unit length of 15.6 dB/cm with a compact spiral waveguide of 5.3 mm length and $ \sim $0.06 mm$ ^{2} $ footprint. The erbium-doped integrated lithium niobate spiral waveguide amplifiers would pave the way in the PICs of the lithium niobate platform, especially in achieving efficient integration of active and passive devices on a lithium niobate thin film, which will make full use of its excellent physical properties such as remarkable photoacoustic, electro-optic, and piezoelectric characteristics.
\end{abstract}

\pacs{}

\maketitle 

	
Integration is of great significance in device miniaturization and improving energy efficiency \cite{bradley2011erbium}. Photonic integrated circuits as one of important goals for the development of photonics have attracted enormous attentions and became one of the most invetigated research fields. Lithium niobate (LN) as an emrging integrated photonic platform material is widely used in optical and microwave fields, due to its rich properties as wide transparent wavelength range, excellent electro-optic, acousto-optic characteristics and large second-order nonlinear susceptibility \cite{nikogosyan2006nonlinear}. In particular, with the lithium niobate on insulator (LNOI) commercializing, more compact and low-cost photonic devices with high-performance can be achieved on LNOI, which is very improtant for the Photonic integrated circuits, especially for the development of a large number of integrated functional devices  \cite{boes2018status,lin2020advances,2017Monolithic}.
Many on-chip optical devices based on LNOI have been demonstrated, such as efficient frequency convetors \cite{luo2018highly,ye2020sum,ge2018broadband,lin2016phase,lin2019broadband}, electro-optical moudulators \cite{xu2020high,li2020lithium,wang2018integrated,pan2020first}, frequency comb source \cite{wang2019monolithic,fang2019efficient,2019Self}.
However, for a complete photonic integrated circuits, integrated waveguide amplifiers and light sources are essential elements to realize on-chip various functionalities. Due to the LN is not a efficient gain medium, we can not achieve waveguide amplifier and lasing directly on a LNOI. In order to overcome this shortage, the most straightforward approches is to doping rare earth ions into the LNOI, which is similar to the erbium doped fiber amplifier. Actually, erbium-doped laser and waveguide amplifeir have been realized in many host materials and show great potential for integrated waveguide amplifiers and light sources  \cite{min2004erbium,agazzi2013energy,mu2020high,ronn2020erbium,vazquez2014erbium,ronn2019ultra}. The on-chip whispering gallery mode lasers based on the LNOI also have been demonstrated, recently \cite{liu2021chip,wang2021chip,yin2021electro,luo2021chip,2021Microdisk}. But, there are still less research works about the on-chip ampilifiers based on the LNOI. It's worth noting that a efficient waveguide amplifier based on erbium-doped thin film LN has been realized recently \cite{chen2021efficient}. 
However, this straight waveguide amplifier with a length of 5-mm is still too long to realize a more compact on-chip photonic integration, especially for a chip with a large number of functional devices.

Here, we demonstrated a spiral waveguide amplifier with a maximum net gain of 8.3 dB on a 1-mol\% erbium-doped LNOI, with a 15.6 dB/cm net gain per unit length. The spiral waveguide amplifiers with total 5.3-mm-long and $ \sim $0.06 mm$ ^{2} $ of footprint based on the CMOS compatible process, presents strong light confinement for the signal and pump light and have great significance on photonic integrated circuits. We shows that it is possible to achieve efficient on-chip amplifiers with a small footprint spiral waveguides, which would pave the way for the active and passive photonic devices integration of various functionalities on the erbium-doped LNOI platform.

   \begin{figure}[]
  	\centering
  	\includegraphics[width=6cm]{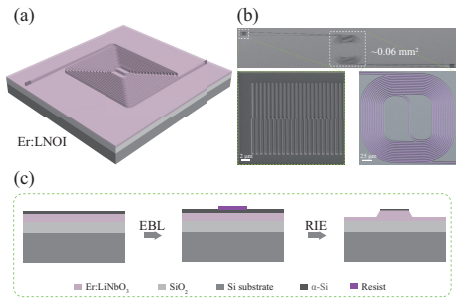}
  	\caption{Schematic of the spiral waveguide amplifiers and its breif fabrication processes. (a) A spiral waveguide amplifier on a erbium-doped LNOI. (b) Scanning electron microscope (SEM) images of the whole spiral waveguide amplifier, and the magnified  coupling grating and spiral waveguide with false-color. (c) The brief fabrication processes of spiral waveguide amplifiers. EBL: electron
  		beam lithography,  RIE: reactive ion etching.}
  \end{figure}

  The spiral waveguide amplifiers were fabricated on a 1-mol $ \% $ Z-cut erbium-doped LNOI, with 600-nm-thick erbium-doped lithium niobate (LN), 2-$ \mu $m-thick silica (SiO$ _{2} $) and 500-$ \mu m $-thick silicon (Si) substrate [Fig. 1(a)]. Here, we select the lithium niobate (LiNbO$_{3} $) as a host materials due to its excellent physical properties compared to orther materials. In order to obtain uniform erboum ions doping concentration and achieve a better gain effect, we doped erbium ions into LN during the crystal growth processes \cite{liu2021chip}.
  
  Figure 1(b) shows the scanning electron microscope (SEM) image of the spiral waveguide amplifier. The total length of the spiral waveguide amplifier is 5.3 mm with a minimum radius of 25 $ \mu$m in the bend part of the spiral waveguides. The footprint of the whole spiral waveguide is $ \sim $0.06 mm$ ^{2} $, which is the smallest among all erbium-doped LNOI waveguide amplifiers \cite{chen2021efficient,zhou2021chip,luo2021chipamplifier}, to the best of our knowledge. Since it will be more suitable for the compact on-chip integration. Insets in Fig. 1(b) are the magnified grating and spiral waveguides part SEM image with false color. In order to obtain high coupling efficiency for pump and signal light. The coupling grating is designed to be two part with periods as 900nm for 1550 nm (Top, duty cycle 0.33) and 1.05 $\mu$m for 980 nm (bottom, duty cycle 0.46). 
  
  The brief fabrication processes of spiral waveguide amplifiers are shown in Fig. 1 (c). Mainly fabrication details including five steps: (1) a 600-nm thick amorphous silicon was deposited as a hard etching mask, (2) a layer of resist was spin-coated onto the Er:LNOI, (3) the spiral waveguide structure was patterned via electron-beam lithography (EBL), (4) the mask layer patterns was transferred to the Er:LN layer with an Ar$ ^{+} $ plasma etching process and (5) the residual mask was removed by wet etching. 
  
  \begin{figure}[]
  	\centering
  	\includegraphics[width=6cm]{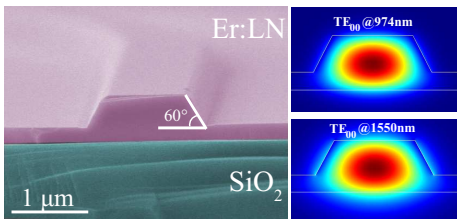}
  	\caption{ The false-color SEM image of the fabricated spiral waveguide cross-section and simulated electric field distributions of the TE$_{00}$ modes at 1550 nm and 974 nm. The top width of spiral waveguide is 1 $ \mu $m, with the thickness around 350 nm and a sidewall angle of $\sim$ $60^{\circ}$}
  	\label{fig:false-color}
  \end{figure}
  
  Figure 2 shows the cross sections of the spiral waveguide (left) and simulated electric field distributions of the TE$_{00}$ modes (right) at pump and signal wavelength. The fabricated spiral waveguides have a top width of 1 $\mu $m, a thickness of 350-nm, and a sidewall angle of $\sim$ $60^{\circ}$, shown as the waveguide cross-section SEM image. From the electric field distribution we find that the micron-scale waveguides can support strong light confinement. The transverse electric (TE) modes are selected both for the pump (980 nm) and signal (C-band) wavelength. 
	
	Figure (3) shows the experimental setup for the characterization of the sprial waveguide amplifier net gain. The spiral waveguide amplifier was pumped by two 980-nm laser sources (LR-MFJ-980, actual output wavelength at 974 nm) from both the input and output sides. A continuous-wave (CW) telecom tunable laser (New Focus TLB-6728, linewidth < 200 kHz, 1520-1570 nm) was combined by a  wavelength division multiplexer (WDM) and coupled into the spiral waveguide from one side. The in-line polarization controllers were used to make the TE polarization light coupling into the device. Then, the output light was collected by a optical spectrum analyzer (OSA) from the 1550 port of the WDM. The top inset shows the photograph of the spiral waveguide amplifier pumped by 974 nm light with strong up-conversion induced green photoluminescence. The bottom inset
	in Fig. 3 shows the  optical microscope image of the small footprint spiral waveguide amplifier comparing with a single-mode optical fiber (Coning SMF-28).

	\begin{figure*}[]
		\centering
		\includegraphics[width=16cm]{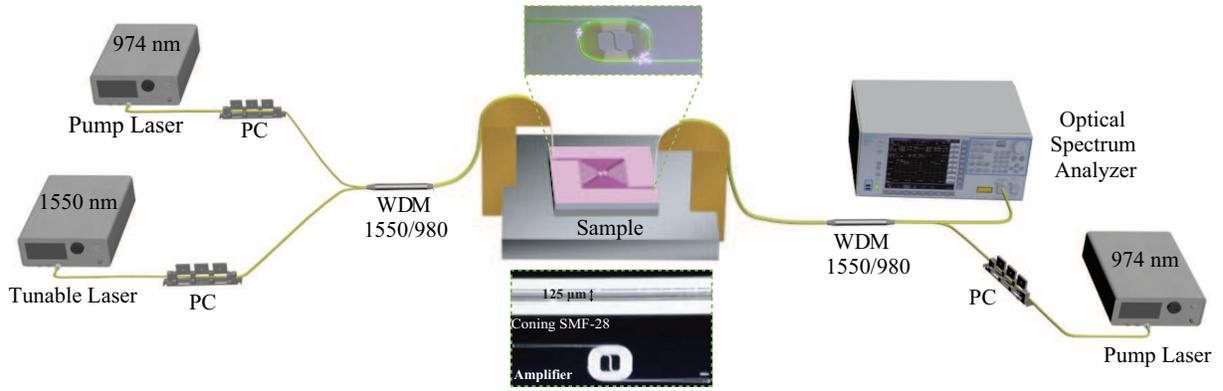}
		\caption{ The net gain measurement setup. Insets are the microscope photograph of the spiral waveguide amplifiers pumped by 974 nm light with strong up-conversion induced green photoluminescence (Top) and the optical microscope image of the spiral waveguide amplifier comparing with a single-mode optical fiber (Coning SMF-28) (bottom).}
		\label{fig:false-color}
	\end{figure*}
	
	We characterized the propagation losses of the spiral waveguide amplifier at the signal wavelength (1550 nm) by a microring. Figure (4) shows the SEM image of the microring with the top width of 1 $\mu $m and radius of 30 $\mu $m, which exhibits the loaded and intrinsic Q factor of  $ 5 \times 10^{4} $ and  $ 5.4 \times 10^{4} $, respectively. We can estimate the propagation loss $ \alpha $ of the spiral waveguide based on the equation $\alpha=2\pi n_{eff}/\lambda_{0}Q_{I}$, where $ n_{eff} $ is the effective index of the waveguide and $ \lambda_{0} $ is the target wavelength.
	The calculated propagation loss at 1550 nm is $ \sim $ 6.86dB/cm, which lead to a loss about 3.64 dB for our 5.3-mm spiral waveguide amplifiers. We also measured the coupling losess of our coupling grating at 980 nm and 1550 nm, which are 16 dB and 13.4 dB, respectively. The launched pump and signal powers into the spiral waveguide have been calibrated by using the above measurement results.
	
	\begin{figure}[]
		\centering
		\includegraphics[width=7cm]{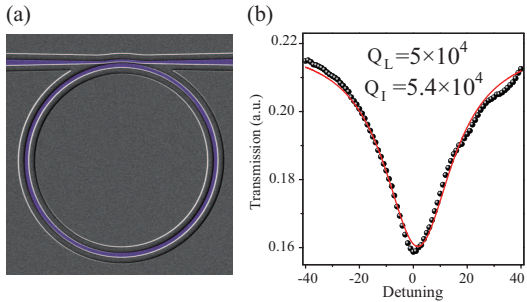}
		\caption{Propagation loss measurement with a micro-ring. (a) The false-color SEM image of microring with the radius of 30 $ \mu  $m. (b) The Lorenztian fitting exhibiting the microring with a loaded Q factor of  Q$_{L} = 5 \times 10^{4} $ and intrinsic Q factor of Q$_{I}=5.4 \times 10^{4} $.}
		\label{fig:false-color}
	\end{figure}

	Figure 5(a) presents the measured signal spectra at 1530 nm with the increasing pump power, which shows the apparently signal enhancement. Figrure 5(b) is the amplifier net gain as a function of the launched pump power with different signal powers at 1530 nm. As expected, the spiral waveguide optical gain increases rapidly at the small pump powers. Then, we can observed the gain approaching saturation with the launched pump power increasing around 10 mW. A maximum net internal gain of 8.3 dB is achieved with the signal power at -10.7dBm and pump power at 10.78 mW, which corresponding to a net gain per unit length of $ \sim $ 15.6 dB/cm. It is  higher than other erbium-doped LNOI \cite{chen2021efficient} and bulk LN \cite{2013Optical,2017Efficient}. What's more, a small gain saturation is also found when the launched signal power increasing, shown as Fig. 5(c). The net gain at orther wavelengths of telecommubication bands is characterized as shown in Fig. 5(d), with a launched signal power at -10.7 dbm. The pink area of Fig. 5(d) shows the net gain over 3 dB and we can find that the spiral waveguide amplifier exhibits a net gain bandwidth of 1530-1570 nm. These results show the potential for on-chip integration with high optical gain amplifier in erbium-doped LNOI platform.
	
	\begin{figure}[]
		\centering
		\includegraphics[width=\linewidth]{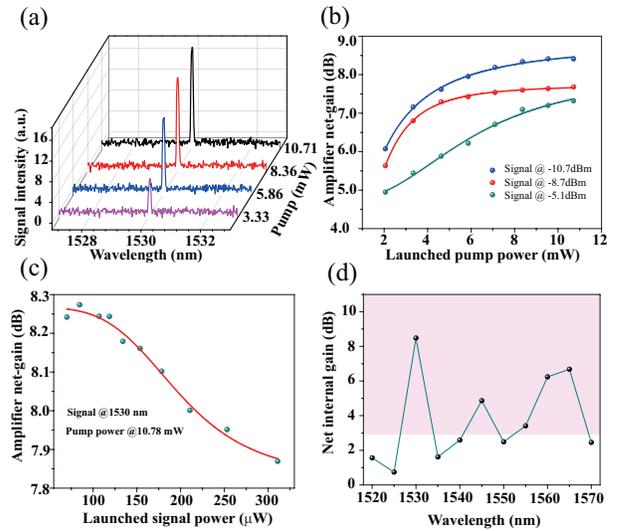}
		\caption{Gain characterization of spiral waveguide amplifiers. (a) Measured signal spectra as a function of launched pump power at 1530 nm. (b) Measured  net internal gain at different launched signal powers (1530 nm) as a function of launched pump power. (c) Measured net internal gain as a function of launched 1530 nm signal powers with a fixed pump powers at 10.78 mW. (d) Net internal gain as a function of signal wavelength with a fixed signal and pump power at -10.7 dBm and 10.78 mW, respectively. }
		\label{fig:false-color}
	\end{figure}

 In conclusion, we fabricated high-gain optical spiral waveguide amplifiers with total 5.3-mm-long and $ \sim $0.06 mm$^{2} $ of areas on a 1-mol\% erbium-doped LNOI. A maximum internal net gain of 8.3 dB at 1530 nm and a broad gain band (1530-1570 nm) have been demostrated. A maximum net gain per unit length can reach up to 15.6 dB/cm. The strong confinement to the pump and signal light, small footprint and relative high signal enhancement of the spiral waveguide amplifier are of great significance for the LN on-chip photonic integrated circuits, which would pave the way in the photonic integrated circuits of lithium niobate platform or the hybrid integration.

It is worth noting that another two erbium-doped waveguide amplifiers works were posted on arXiv \cite{zhou2021chip, luo2021chipamplifier}, during the preparation of this article. Comparing to these two works, our spiral waveguide amplifiers have a narrower top width and smallest footprint, which means the spiral waveguide amplifiers are integrated and can support more compact on-chip integration. Spiral waveguide amplifiers should be better for a large number of photonic devices integration.

This work was supported by the National Key R $\&$ D Program of China (Grant Nos. 2019YFB2203501, and 2017YFA0303701, 2018YFA0306301), the National Natural Science Foundation of China (Grant Nos. 91950107, 11734011), Shanghai Municipal Science and Technology Major Project
(2019SHZDZX01-ZX06), and SJTU No. 21X010200828.	

 We thank Dr. Hao Li for providing ICP-RIE etching help for this device.
 
[$ ^{\ast} $] These authors contributed equally to this Letter.		

\bibliography{References}

\end{document}